\documentclass[12pt]{article}

\normalbaselines \topmargin -0.25 truein \oddsidemargin 0.30
truein \evensidemargin 0.30 truein 

\begin{document}

\newcommand{\nablab}{{\mathop {\rule{0pt}{0pt}{\nabla}}\limits^{\bot}}\rule{0pt}{0pt}}

\vspace{15mm}

\centerline{\bf  AXION ELECTRODYNAMICS AND DARK MATTER
FINGERPRINTS}

\centerline{\bf IN THE TERRESTRIAL MAGNETIC AND ELECTRIC FIELDS}

\vskip 0.3cm \centerline{\bf A.B. Balakin\footnote{e-mail:
Alexander.Balakin@ksu.ru} }

\vskip 0.3cm

\centerline{\it Department of General Relativity and Gravitation,
 Kazan Federal University,} \centerline{\it
Kremlevskaya str., 18, Kazan, 420008, Russia}

\vskip 0.3cm \centerline{and}
\vskip
0.3cm \centerline{\bf L.V. Grunskaya\footnote{e-mail: grunsk@vlsu.ru}
}
\vskip 0.3cm
\centerline{\it Department of General and Applied Physics, Vladimir State University,}
\centerline{\it Gorky str., 87, Vladimir, 600000, Russia}

\vspace{5mm}
\noindent
{\bf Abstract}

\noindent {\small We consider mathematical aspects of the axion
electrodynamics in application to the problem of evolution of
geomagnetic and terrestrial electric fields, which are coupled by
relic axions born in the early Universe and (hypothetically)
forming now the cold dark matter. We find axionic analogs of the
Debye potentials, well-known in the standard Faraday - Maxwell
electrodynamics, and discuss exact solutions to the equations of
the axion electrodynamics describing the state of  axionically
coupled electric and magnetic fields in a spherical resonator
Earth-Ionosphere. We focus on the properties of the specific
electric and magnetic oscillations, which appeared as a result of
the axion-photon coupling in the dark matter environment. We
indicate such electric and magnetic field configurations as
longitudinal electro-magnetic clusters.}

\vspace{5mm}


\newpage

\section{Introduction}

\subsection{On the physical aspects of the problem of the axion - photon coupling}

The term {\it dark matter axions}, which came into general use
during the last decade, reflects a synthesis of two trends based
on ideas of modern astrophysics and cosmology on the one hand, and
of high-energy particle physics on the other hand. Dark matter,
which neither emits nor scatters electromagnetic radiation, is
considered to be one of the main cosmic substrates accumulating
about 23\% of the Universe energy. Mass density distribution of
the dark matter is now well-studied due to observations and
theoretical modeling (see, e.g., \cite{1}-\cite{4} for details,
review and references). The origin of the dark matter is not yet
established, but there are several hypotheses about its nature.
The most attractive is the hypothesis that the dark matter is
formed by massive  pseudo-Goldstone bosons (axions) belonging to
the class of WIMPs (Weakly Interacting Massive Particles). The
story of axions, which were predicted by Peccei and Quinn
\cite{Peccei} and introduced into the high-energy physics as new
light bosons by Weinberg \cite{Weinberg} and Wilczek
\cite{Wilczek0}, is well-described: one can find reviews and
references, e.g., in \cite{8}-\cite{14}.

We are interested in studying of the axion-photon coupling; more
precisely, we consider the response of the photon system to the
action of the axions and possible experimental consequences of
such action. First of all, it is worth reminding what type of
axions we are going to consider in this paper. It is well-known
that axions can be generated if there exist strong magnetic field,
strong electric field, and they are not orthogonal to one another.
Such axion sources can exist in the Solar plasma, in the
magnetosphere of neutron stars, and in laboratories on the Earth.
Other mechanisms of the axion production are predicted to occur in
the early Universe; the axions born in the early Universe form now
the unique cosmological substrate indicated as relic (or
primordial, or background) axions; these are the cosmological dark
matter axions, and their density distribution depends on time.
Since the axions are massive particles, their gravitational
attraction forms non-homogeneous configurations; for instance, the
density distribution in the galactic halos is stationary but
depends on the distance from the galactic center. In other words
we live (hypothetically) in an axionic environment and have a
right to pose a question: how serious the influence of the axions
on the terrestrial electrodynamic systems is.

Wave propagation in the axion electrodynamics is studied in detail
for two cases: first, when one deals with running plane waves,
second, when one uses the approximation of geometrical optics,
i.e., the problem of dispersion relations can be effectively
reduced to the analysis of the Fresnel equation (see, e.g.,
\cite{Itin1,Itin2,plane}). It was shown that the axion-photon
interactions induce the effect of polarization rotation, if the
electromagnetic waves travel through the axion system (see, e.g.,
\cite{HO1}- \cite{Ni11} for details). The inertial-gradient extension of
the Einstein-Maxwell-axion model \cite{BT} predicts that the
optical activity induced by relic cosmological axions should be a
nonstationary process and should keep information about the rate
of the Universe expansion. Nonminimal extension of the
Einstein-Maxwell-axion model \cite{BaWTN} allows us to describe
the birefringent wave propagation. Minimal two-parameter
gradient-type extension of the Einstein-Maxwell-axion model
\cite{BBT} also predicts the existence of birefringence, as well
as, could explain abnormal growth of the number of axions in the
early Universe.

In this paper we focus on the electrodynamic effects, which can
occur in the terrestrial electric and magnetic fields under the
influence of the relic dark matter axions. Relic axion
distribution is considered to be homogeneous but time dependent.
We work here within the Weinberg - Wilczek - Ni version of the
axion electrodynamics, and focus on the axionically coupled
electromagnetic oscillations in a spherical resonator. The
obtained results are new, they can have numerous applications and
are planned to be generalized in the framework of nonminimal and
gradient-type extensions of the Einstein - Maxwell - axion model.

\subsection{On the mathematical aspects of the problem of the axion - photon coupling}

The axion-photon interaction can be described both in terms of
high-energy physics and in terms of macroscopic axion
electrodynamics. We analyze the problem in the framework of axion
electrodynamics in the Weinberg - Wilczek - Ni (WWN) version
\cite{Weinberg,Wilczek0,Ni,Wilczek87}. The WWN model is the
particular case of the Carroll - Field - Jackiw (CFJ) model
\cite{CFJ}, based on the Chern - Simons ansatz \cite{CS}. These
models are often discussed in context of the Equivalence Principle
and  Lorentz violation (see, e.g., \cite{LV1} - \cite{LV5} for
review and references). Two models: CFJ and WWN, are known to be
effectively coinciding, when the pseudovector $p_i$ appearing in
the first model is proportional to the four-gradient of the
pseudoscalar field $\phi$, on which the WWN model is based (e.g.,
$p_i = \nabla_i \phi$). Let us mention that in the cosmological
context, when pseudoscalar field is the function of time only,
$\phi(t)$, and the pseudovector $p_i$ is time-like and has only
one component, $p_i=\delta_i^0 \psi(t)$, both electrodynamic
extensions: WWN and CFJ, - give the same results. It gave us a
supplementary motivation to use the WWN version of the axion
electrodynamics in this paper.

Generally, the axion electrodynamics in the Weinberg - Wilczek -
Ni version is more sophisticated from the mathematical point of
view than the Faraday - Maxwell theory. The main extension is that
the axion electrodynamics includes additional equation for the
pseudoscalar (axion) field $\phi$, which, in its turn, enters the
equations for the electromagnetic field strength $F_{ik}$ (the
Maxwell tensor), as well as, the constitutive equations. The total
self-consistent system of master equations is nonlinear. The set
of boundary and initial conditions is extended accordingly. We are
interested to solve the master equations of the axion
electrodynamics, which guide the evolution of the magnetic and
electric field in the spherical resonator "Earth-Ionosphere". In
the framework of the Faraday - Maxwell electrodynamics this
problem is solved and discussed in detail, in particular, in terms
of Hertz and Debye potentials \cite{book}. These results are classical  and
they were used in numerous applications, in particular, in radio-
and tele-communication, navigation, magnetic storm predictions, etc.
When the axion-photon coupling is taken into account, these
results require serious modifications, since in this case the
electromagnetic wave propagate in chiral (quasi)medium. For
instance, in classical terrestrial electrodynamics the so-called
$E$-waves and $H$-waves are introduced \cite{book}, for which the
radial components of the magnetic and electric fields, respectively, are
equal to zero. From the mathematical point of view, such decomposition
of the total electromagnetic field essentially facilitates the
analysis of the problem. In the axion electrodynamics the
introduction of pure $E$- wave or pure $H$- wave is not possible,
since the axionic environment rotates the polarization of the
electromagnetic waves.

Our goal is to find some analogs of the Debye potentials for the
model, which takes into account the axion - photon coupling.
Specific feature of our approach is that the pseudoscalar field is
considered to be formed by relic cosmological axions. This means,
first, that the axion field $\phi$ is specified, and thus the
corresponding evolutionary equation for the pseudoscalar field is
decoupled from the total set of equations of the axion
electrodynamics. This approximation is well-motivated, since the
density of dark matter axions is much higher than the number of
axions  produced by terrestrial electromagnetic field itself. The
second specific detail, which also is well-motivated in the
context of relic cosmological axions, is that the pseudoscalar
field $\phi(t)$ evolves with time much more slowly than the
typical terrestrial electric and magnetic fields oscillate. More
precisely, in our model we consider the time derivative of the
pseudoscalar field $\dot{\phi}$ as a constant, and introduce
appropriate quantity $q$, which is directly connected with actual
mass density of the axionic dark matter in the vicinity of the
Earth.

In other words, we focus on a typical problem of mathematical
physics, which came from the theory of evolution of the
terrestrial electric and magnetic fields, the axion - photon
coupling being the new key element of theory.

\subsection{Organization of the paper}

The paper is organized as follows. In Section 2 we formulate the
model equations of axion electrodynamics for the case, when the
pseudoscalar field is formed by relic cosmological axions born in
the early Universe and is revealed now as axionic dark matter. In
Subsection 2.3 we consider two illustrations: one exact  and one
approximate solutions to these equations, which show that relic
axions can form specific electrodynamic configurations with
parallel coupled electric and magnetic fields. In Section 3 we
introduce the so-called truncated ($\varphi$-independent) model
and represent the reduced model equations in spherical
coordinates. Section 4 is devoted to the analysis of static
solutions to the basic equations: we discuss the axion
electrostatics in Subsection 4.1, the axion magnetostatics in
Subsection 4.2. In Subsection 4.3 we reduce the obtained results
to the dipole-type model of the terrestrial magnetic field, and
consider some properties of this model in Subsection 4.4. In
Section 5 we analyze time-dependent solutions to the basic
equations of the truncated ($\varphi$-independent) model in the
spherical resonator: in Subsection 5.1 we obtain eigen-functions
for the boundary-value problem, frequencies and amplitudes of the
axionically coupled modes; in Subsection 5.2 we solve the problem
for the case, when the azimuthal electric current induces forced
oscillations. In Section 6 we focus on the description of
axionically coupled electric and magnetic oscillations near the
Earth surface. Section 7 is devoted to discussion of the obtained
results.

\section{Axion-photon coupling in the axionic dark matter environment}

\subsection{The model}

Let us start with the action functional
\begin{equation}
S {=} \frac{\hbar}{c} \int d^4 x \sqrt{{-}g} \left\{\frac
{R{+}2\Lambda}{\kappa}{+}\frac{1}{2}F_{mn}F^{mn}
{+}\frac{1}{2}\phi F^{*}_{mn} F^{mn} {-} \Psi_{0}^2\left[
\nabla_m\phi \nabla^m\phi {-}V(\phi^2) \right]\right\} .
\label{action02}
\end{equation}
Here, $g$ is the determinant of the metric tensor $g_{ik}$,
$\nabla_{m}$ is the covariant derivative, $R$ is the Ricci scalar,
$\kappa \equiv \frac{8\pi G}{c^4}$, $G$ is the gravitational
Newtonian coupling constant, $\Lambda$ is the cosmological
constant, $\hbar$ is the Planck constant and $c$ is the speed of
light in vacuum. The Maxwell tensor $F_{mn}$ is given by
\begin{equation}
F_{mn} \equiv \nabla_m A_{n} - \nabla_n A_{m} \,, \quad \nabla_{k}
F^{*ik} =0 \,, \label{maxtensor}
\end{equation}
where $A_m$ is a potential four-vector of the macroscopic electromagnetic field; $F^{*mn}
\equiv \frac{1}{2} \epsilon^{mnpq}F_{pq}$ is the tensor dual to
$F_{pq}$; $\epsilon^{mnpq} \equiv \frac{1}{\sqrt{-g}} E^{mnpq}$ is
the Levi-Civita tensor, $E^{mnpq}$ is the absolutely antisymmetric
Levi-Civita symbol with $E^{0123}{=}1$.

The first term in (\ref{action02}) is the Hilbert-Einstein
Lagrangian for the gravity field with the account of the
cosmological constant. The second term $\frac{1}{2}F_{mn} F^{mn}$
relates to the standard linear electrodynamics in vacuum. The term
$\frac{1}{2}\phi F^{*}_{mn} F^{mn}$ in (\ref{action02}) describes
the pseudoscalar-photon interaction \cite{Ni}. The symbol $\phi$
stands for an effective pseudoscalar field; this quantity is
dimensionless providing the second and third terms in the integral
to be of the same dimensionality. The macroscopic effective axion
field, $\Phi$, is considered to be proportional to this quantity
$\Phi {=} \Psi_0 \phi$, and the constant $\Psi_0$ is the subject
of a special discussion. The parameter $\Psi_{0}^2$ has the
dimensionality of a force,  and its estimation depends
substantially on the model of the axion coupling, on the value of
the axion-photon coupling constant $g_{({\rm A}\gamma \gamma)}$,
on the individual axion mass $m_{({\rm A})}$, on the number of
axions in the unit volume ${\cal N}_{({\rm A})}$, etc. (see, e.g.,
\cite{10,n1}). The function $V(\phi^2)$ describes the potential of
the pseudoscalar field; in the simplest case we use the quadratic
potential $V {=} \mu^2 \phi^2$, where $\mu {=} \frac{m_{({\rm
A})}c}{\hbar}$ has the dimensionality of inverse length.

Below we assume that the gravitational background is given, the
space-time being of the Friedmann - Lema\^itre - Robertson -
Walker (FLRW) type
\begin{equation}
ds^2 = a^2(x^0)\left[ (dx^0)^2 - dl^2 \right] \,, \label{metric1}
\end{equation}
with the scale factor $a$ and the Hubble function
\begin{equation}
H = \frac{1}{a^2} \frac{d a}{d x^0} = \frac{\dot{a}}{a} \,.
\label{ax19}
\end{equation}
Here and below the dot denotes the derivative with respect to time $t$, which is connected with $x^0$ by the differential relation
\begin{equation}
a(x^0) \ dx^0 = c dt \,. \label{metric7}
\end{equation}
The metric in the three-space is assumed to be represented in the spherical coordinates
\begin{equation}
dl^2 = dr^2 + r^2(d\theta^2 + \sin^2{\theta} d \varphi^2)\,,
\label{metric}
\end{equation}
and we use the formula $\sqrt{-g}= a^4 \ r^2 \sin{\theta}$. This
means that we do not consider backreaction of the electromagnetic
field on the gravitational field; moreover, we neglect the direct
influence of the terrestrial gravity field on the electric and
geomagnetic fields in comparison with the influence of the relic
cosmological axions.

The effective macroscopic pseudoscalar field, appearing in the
action functional (\ref{action02}), satisfies the equation
\begin{equation}
\nabla_k \nabla^k \phi + \phi V'(\phi^2) = -
\frac{1}{4\Psi_{0}^2} F^{*}_{mn}F^{mn} \,,
\label{ax1}
\end{equation}
however, we assume that the number of axions produced by the
macroscopic electromagnetic field $F_{ik}$ is much less than the
number of relic (primordial) axions created in the early Universe.
This means that we neglect the electromagnetic source in the
right-hand side of the equation (\ref{ax1}) and consider the
function $\phi(t)$ to satisfy the decoupled equation
\begin{equation}
\ddot{\phi} + 3H \dot{\phi} +  \mu^2 c^2 \phi = 0 \,.
\label{ax2}
\end{equation}
The last and very important element of this macroscopic model
comprises the relation between the effective pseudoscalar field
$\phi$ and the dark matter state functions, the energy density
$W(t)$ and pressure $P(t)$. These relations appear when we put
equal the effective stress-energy tensor of the macroscopic
pseudoscalar field on the one hand, and the phenomenological
stress-energy tensor of the dark matter fluid, on the other hand
(see, e.g., \cite{ax11,BBT} for details); they have the following
form
\begin{equation}
\frac{1}{2}\Psi^2_0 \left[\frac{q^2}{a^2} + \mu^2 \phi^2 \right] =
W \,, \label{cl15}
\end{equation}
\begin{equation}
\frac{1}{2}\Psi^2_0 \left[\frac{q^2}{a^2} - \mu^2 \phi^2 \right] =
P \,. \label{cl17}
\end{equation}
The quantity $q$ is defined as
\begin{equation}
q \equiv \frac{d \phi}{d x^0} = \frac{a}{c} \ \dot{\phi} \,,
\label{q1}
\end{equation}
and can be readily found as
\begin{equation}
q = \pm \frac{a}{\Psi_0} \sqrt{W+P} \,. \label{cl18}
\end{equation}
For the cold dark matter $P \to 0$ and  $W \to \rho c^2$,
thus
\begin{equation}
q \to  \pm \frac{c a}{\Psi_0} \sqrt{\rho_{({\rm DM})}} \,,
\label{cl154}
\end{equation}
where $\rho_{({\rm DM})}$ is the mass density of the cold dark matter.

\subsection{Electrodynamic equations}

The equations of the axion electrodynamics have the following form
\begin{equation}
\nabla_k \left[F^{ik} {+} \phi  F^{*ik} \right] = - \frac{4
\pi}{c} J^i \,, \label{eld1}
\end{equation}
where $J^i$ is the electric current four-vector. Taking into account
the equations (\ref{maxtensor}) one can rewrite these equations as follows:
\begin{equation}
\nabla_k F^{ik} = - \frac{4 \pi}{c} \left[J^i + J^i_{({\rm A})}
\right]\,, \label{0eld1}
\end{equation}
where
\begin{equation}
\frac{4 \pi}{c} J^i_{({\rm A})} \equiv F^{*ik} \nabla_k \phi  \,.
\label{0eld102}
\end{equation}
When the electric current four-vector $J^i$ describes the standard
conductivity, we can write
\begin{equation}
J^i = \sigma \ E^i = \sigma \ F^{ik} U_k =
\frac{1}{2}(g^{im}g^{kn}-g^{in}g^{km}) \ \sigma \ U_k F_{mn}\,,
\label{0eld101}
\end{equation}
where $U^k$ is the four-vector of the macroscopic velocity of the system as a whole. Similarly, we can write
\begin{equation}
\frac{4 \pi}{c} J^i_{({\rm A})} =  \frac{1}{2} \epsilon^{ikmn}
\nabla_k \phi F_{mn}  \,, \label{20eld102}
\end{equation}
and link this term with axionic conductivity. Clearly, the
four-pseudovector $\frac{c}{4\pi}\nabla_k \phi$ replaces the
four-vector $\sigma U_k$, when the axionic conductivity is
discussed.

\subsection{Prologue: 3-dimensional representation of the model and spatially homogeneous solutions}

In 1987 Wilczek (see the paper \cite{Wilczek87}) considered the
application of axion electrodynamics to the problem of spherically
symmetric static dyons. The dyon-type solution  was obtained using
two ingredients: a magnetic monopole plus axion field. In this
case the static axion field produces static radial electric field
using the radial magnetic field of the monopole. In fact, Wilczek
presented the first (static) example of the longitudinal
electro-magnetic cluster, since the static electric and magnetic
fields are parallel in this spherically symmetric configuration.
Now we illustrate the idea of longitudinal clusters by the
example, in which the electric, magnetic and axion fields are
spatially homogeneous, but depend on time.

Equations of the axion electrodynamics can be written in the
three-vectorial form as follows:
\begin{equation}
{\rm rot} \vec{E} = - \frac {1}{c} \frac{\partial}{\partial t}
\vec{B} \,, \quad {\rm div} \vec{E} = 4\pi \rho - \vec{B} \cdot
\vec{\nabla} \phi \,, \label{ww1}
\end{equation}
\begin{equation}
{\rm div} \vec{B} = 0 \,, \quad {\rm rot} \vec{B} = \frac {1}{c}
\frac{\partial}{\partial t} \vec{E} + \frac {4\pi}{c} \vec{J} +
\frac {1}{c} \vec{B}\frac{\partial}{\partial t} \phi -
\vec{E} \times \vec{\nabla} \phi  \,. \label{ww2}
\end{equation}
We assume here that the pseudoscalar (axion) field depends on time
only, i.e., $\phi{=}\phi(t)$, and $\vec{\nabla} \phi {=}0$. Also,
we assume that $\rho {=} 0$ and $\vec{J} {=} 0$, i.e., we study
electrodynamic effects in the regions out of the electric charges.
Then there exists the following exact constant solution: the
magnetic field is constant, i.e., $\vec{B} {=} \vec{B}_0 {=}
const$, and the electric field is homogeneous, i.e., $\vec{E}{=}
\vec{E}(t)$. Indeed, all the equations (\ref{ww1}) and (\ref{ww2})
are satisfied, if
\begin{equation}
0 = \frac{d}{d t} \vec{E} +  \vec{B} \frac{d}{d t} \phi \,.
\label{0ww2}
\end{equation}
Clearly, the exact solution for the electric field reads
\begin{equation}
\vec{E}(t)= \vec{E}(0) - [\phi(t)-\phi(0)] \vec{B}_0 \,.
\label{0ww21}
\end{equation}
This means that in the presence of constant magnetic field
in addition to the initial arbitrary directed electric field $\vec{E}(0)$
the electric field collinear to this magnetic field appears due to the
influence of the evolving system of relic axions.

Now, let us consider an illustrative approximate model, in which
the magnetic field is not constant, $\vec{B}{=}\vec{B}(t)$, but it
varies with time very slowly. Such situation is typical for the
problem of infra-low-frequency variations of the terrestrial
electric and magnetic fields. For instance, the variations of the
geomagnetic field on the frequency $\omega/2\pi {=} 10^{{-}3} Hz$
have the effective wavelength of the order $\lambda {=} 300\cdot
10^{6}$ km, which is the distance to the Sun and back. Thus, the
right-hand side of the first equation in (\ref{ww1}) can be
considered negligible, and the electric field, again, can be
treated as homogeneous. Then, one obtains from (\ref{ww2})
\begin{equation}
\vec{E}(t) = \vec{E}(0) -  c \int_0^t d\tau q(\tau) \vec{B}(\tau)
\,. \label{0ww29}
\end{equation}
In comparison with the infra-low-frequency variations of the geomagnetic field
the function $q(t)$ behaves as a constant, and we can use the simplified formula
\begin{equation}
\vec{B(t)} = \vec{B}_0 + \delta \vec{B} \cdot \cos{\omega t} \ \rightarrow \ \vec{E}(t) = \vec{E}(0) -
[\phi(t)-\phi(0)] \vec{B}_0 - \frac{c q}{\omega} \delta \vec{B} \cdot \sin{\omega t} \,,
\label{0ww3}
\end{equation}
where $\delta \vec{B}$ denotes the amplitude of the geomagnetic
field variation, the quantity $\omega$ being its frequency.
This is
a good illustration of our idea to focus on the analysis of the
electric field variations produced by the geomagnetic field
variations in the relic axion background. Clearly, this electric
field induced by the axions is parallel to the geomagnetic field
variation, that is why we use the term "longitudinal
electro-magnetic cluster" to indicate this specific mode in the
combined electric and magnetic fields variations.

\section{Electrodynamics in the relic axion background}

\subsection{Truncated ($\varphi$-independent) model}

Let us now apply the model to the case of terrestrial electric and
magnetic fields. In fact, the geomagnetic field is
non-homogeneous, and we have to take into account the curvature of
the magnetic field lines when consider global effects in the
Earth's magnetosphere. Below we discuss the non-stationary
electric and magnetic fields, and start with the potentials, which
do not depend on the azimuthal variable $\varphi$, i.e., $A_i {=} A_i(t,r,\theta)$
for $i {=} 0,r,\theta,\varphi$.
This model is truncated, nevertheless, it illustrates all the
important properties of the axion-photon coupling in the Earth's
electrodynamic system.

One can introduce the truncated ($\varphi$-independent) model
using the following approach. Let us remind that the flat
space-time metric admits the existence of the Killing vector
$\xi^i_{(\varphi)} = \delta^i_{\varphi}$ and the Lie derivative of
the metric is equal to zero,
\begin{equation}
\pounds_{\xi_{(\varphi)}} g_{mn} = \xi^l_{(\varphi)}
\frac{\partial }{\partial x^l} g_{mn} + g_{ln}
\frac{\partial}{\partial x^m} \xi^l_{(\varphi)} + g_{ml}
\frac{\partial}{\partial x^n} \xi^l_{(\varphi)}  = 0 \,,
\label{Lie1}
\end{equation}
so in the standard spherical coordinates the metric  does not depend on $\varphi$.
Let us assume that the potential four-vector $A_k$ inherits this symmetry, i.e.,
\begin{equation}
\pounds_{\xi_{(\varphi)}} A_{k} = \xi^l_{(\varphi)} \frac{\partial
}{\partial x^l} A_{k} + A_{l} \frac{\partial}{\partial x^k}
\xi^l_{(\varphi)} = 0 \,.
\label{Lie2}
\end{equation}
This provides the condition $\frac{\partial A_k}{\partial \varphi}
{=} 0$.

\subsection{Master equations}

The equations (\ref{eld1}) can be written as follows.
The first equation (for $i{=}0$)
\begin{equation}
\frac{1}{r^2}\frac{\partial}{\partial r}(r^2 F_{0r}) +
\frac{1}{r^2 \sin{\theta}}\frac{\partial}{\partial \theta}
(\sin{\theta} F_{0\theta}) = 4\pi \rho \,, \label{t2}
\end{equation}
where  $\rho \equiv \frac{a^4}{c} J^0 $, does not contain the quantity $q$. Three equations for $i
{=}r,\theta,\varphi$ give, respectively,
\begin{equation}
\frac{\partial}{\partial x^0} F_{0r} + \frac{1}{r^2
\sin{\theta}}\frac{\partial}{\partial \theta} (\sin{\theta}
F_{r\theta}) = \frac{q}{r^2 \sin{\theta}} F_{\theta \varphi} -
\frac{4\pi}{c} I^{r} \,, \label{t3}
\end{equation}
\begin{equation}
\frac{\partial}{\partial x^0}F_{\theta 0} +
\frac{\partial}{\partial r} F_{r\theta} = \frac{q}{\sin{\theta}}
F_{r \varphi} + \frac{4\pi}{c} r^2 I^{\theta} \,, \label{t4}
\end{equation}
\begin{equation}
\frac{\partial}{\partial x^0}F_{0 \varphi} +
\frac{\partial}{\partial r} F_{\varphi r} +
\sin{\theta}\frac{\partial}{\partial \theta}
\left(\frac{F_{\varphi \theta}}{r^2 \sin{\theta}}\right) = q
\sin{\theta} F_{r \theta} - \frac{4\pi}{c} r^2 I^{\varphi}
\sin^2{\theta} \,. \label{t5}
\end{equation}
Here the following notations have been used
\begin{equation}
I^r = a^4 J^{r} \,, \quad I^{\theta} = a^4 J^{\theta} \,, \quad
I^{\varphi} = a^4 J^{\varphi} \,,
\label{not}
\end{equation}
in these terms the equations (\ref{t2})-(\ref{t5}) do not contain
the scale factor $a(x^0)$ at all, so that the only quantity $q$
reminds that these are the model equations of axion
electrodynamics.

The components of the electric current four-vector are linked by
the charge conservation law
\begin{equation}
c \frac{\partial}{\partial x^0} \rho +
\frac{1}{r^2}\frac{\partial}{\partial r}(r^2 I^r) +
\frac{1}{\sin{\theta}} \frac{\partial}{\partial \theta}
(\sin{\theta \ I^{\theta}}) = 0 \,. \label{t51}
\end{equation}
The second subset of equations in (\ref{maxtensor}) gives one equation of the form
\begin{equation}
\frac{\partial}{\partial \theta} F_{0r} +   \frac{\partial}{\partial r} F_{\theta 0} +   \frac{\partial}{\partial x^{0}} F_{r \theta} =0
\,,
\label{t14}
\end{equation}
which does not include the symbol $\varphi$. Other equations
\begin{equation}
\frac{\partial}{\partial \theta} F_{\varphi r} +   \frac{\partial}{\partial r} F_{\theta \varphi} =0 \,,
\quad \frac{\partial}{\partial \theta} F_{0 \varphi} +   \frac{\partial}{\partial x^{0}} F_{\varphi \theta} =0
\,,
\quad \frac{\partial}{\partial r} F_{\varphi 0} +   \frac{\partial}{\partial x^{0}} F_{r \varphi} =0
\,,
\label{t144}
\end{equation}
contain only two terms for the truncated model under
consideration.

Linearity of the electrodynamic equations allows us to consider
separately static and time-dependent solutions of the model. Let
us start with the static ones.

\section{Static electric and magnetic fields}

Since the axion field changes extremely slow, and we suppose $q
{=}const$, it is convenient to reconstruct the electric and
magnetic fields in terms of two potential components, namely,
$A_0(r,\theta)$ and $A_{\varphi}(r,\theta)$. Let us demonstrate
the procedure and the obtained results for electric and magnetic
fields out of domain of charges distribution.

\subsection{Axion electrostatics}

As usual, we distinguish the coordinate and physical (with subscripts in parentheses) components of the electric field, defined as follows
\begin{equation}
E^r \equiv F^{r0} = - \frac{\partial A_{0}}{\partial r} \equiv E_{({\rm rad})} \,,
\label{ms2r}
\end{equation}
\begin{equation}
E^{\theta} \equiv F^{\theta 0} = - \frac{1}{r^2}\frac{\partial A_{0}}{\partial \theta} \equiv \frac{1}{r}E_{({\rm merid})} \,,
\label{ms2t}
\end{equation}
\begin{equation}
E^{\varphi} \equiv F^{\varphi 0} = - \frac{1}{r^2 \sin^2{\theta}}\frac{\partial A_{0}}{\partial \varphi} \equiv \frac{1}{r \sin{\theta}}E_{({\rm azim})} = 0\,.
\label{ms2p}
\end{equation}
Let us remind the motivation of these definitions.
Since there exist two sets of quantities: the covariant and contravariant components of the electric field three-vector ($E_r, E_{\theta}, E_{\varphi}$ and $E^r, E^{\theta}, E^{\varphi}$, respectively), one needs to introduce the so-called physical components, which are assumed to be measured in experiments. We follow the Synge's concept (see, e.g., \cite{Synge}) and introduce
the tetrad four-vectors
\begin{equation}
\lambda^i_{(0)} = \frac{1}{a} \delta^i_0 \,, \quad \lambda^i_{({\rm rad})} = - \frac{1}{a} \delta^i_{r} \,, \quad \lambda^i_{({\rm merid})} = - \frac{1}{a r} \delta^i_{\theta} \,, \quad  \lambda^i_{({\rm azim})} = - \frac{1}{a r \sin{\theta}} \delta^i_{\varphi}
\,,
\label{tetrad1}
\end{equation}
which are orthogonal one to another and satisfy the normalization conditions
\begin{equation}
g_{ik} \lambda^i_{(0)} \lambda^k_{(0)} {=} 1 \,, \quad g_{ik} \lambda^i_{({\rm rad})} \lambda^k_{({\rm rad})} {=} {-}1 \,, \quad g_{ik}\lambda^i_{({\rm merid})}\lambda^k_{({\rm merid})} {=} {-} 1 \,, \quad  g_{ik}\lambda^i_{({\rm azim})}\lambda^k_{({\rm azim})} {=} {-} 1
\,,
\label{tetrad3}
\end{equation}
when the space-time metric has the form (\ref{metric1}) with (\ref{metric}).
The corresponding  components of the electric field three-vector are defined as
\begin{equation}
E_{({\rm rad})} \equiv \frac{1}{a} E_i \lambda^i_{({\rm rad})} \,, \quad  E_{({\rm merid})} \equiv \frac{1}{a} E_i \lambda^i_{({\rm merid})} \,, \quad E_{({\rm azim})} \equiv \frac{1}{a} E_i \lambda^i_{({\rm azim})}
\,,
\label{tetrad2}
\end{equation}
providing the relationships (\ref{ms2r})-(\ref{ms2p}).
In terms of standard three-dimensional metric (\ref{metric}) these quantities can be formally expressed as
\begin{equation}
E_{({\rm rad})} \equiv \sqrt{|E^r E_r|} \,, \quad E_{({\rm merid})} \equiv \sqrt{|E^{\theta} E_{\theta}|} \,, \quad E_{({\rm azim})} \equiv \sqrt{|E^{\varphi} E_{\varphi}|} \,.
\label{ph}
\end{equation}
The equation (\ref{t2}) with $\rho{=}0$ yields the truncated
Laplace equation
\begin{equation}
\Delta_{(0)} A_0 = 0 \,.
\label{0m0}
\end{equation}
Here and below we use the following truncated Laplace operator for
the sake of convenience:
\begin{equation}
\Delta_{(m)} \equiv \frac{1}{r^2}\frac{\partial}{\partial
r}\left(r^2 \frac{\partial}{\partial r}\right) + \frac{1}{r^2}
\left[\frac{1}{\sin{\theta}} \frac{\partial}{\partial \theta}
\left(\sin{\theta}\frac{\partial}{\partial \theta} \right) -
\frac{m^2}{\sin^2{\theta}}\right] \,. \label{10m0}
\end{equation}
The equation (\ref{0m0}) does not contain the quantity $q$, so,
the corresponding solution is well-known in classical mathematical
physics. E.g., for the domain $r>R$ it has the form
\begin{equation}
A_0(r,\theta) = \sum_{n=0}^{\infty} C_n \left(\frac{R}{r}
\right)^{n+1} P_n(\cos{\theta}) \,, \label{1m01}
\end{equation}
where  $R$ is the Earth radius. We use the standard definitions:
$P_n(x)$ for the Legendre polynomials \cite{Jackson,book} and
$P^{(m)}_n(x) = \left(1{-}x^2\right)^{\frac{m}{2}}
\left(\frac{d}{dx}\right)^m P_n(x)$ for the adjoint Legendre
polynomials. As usual, the coefficients $C_n$ can be found from
boundary conditions (we do not consider this problem here).
Clearly, the static electric field components
\begin{equation}
E_{({\rm rad})} = \sum_{n=0}^{\infty} \frac{(n+1)}{R} C_n \left(\frac{R}{r} \right)^{n+2} P_n(\cos{\theta}) \,,
\label{m01}
\end{equation}
\begin{equation}
E_{({\rm merid})} = \sum_{n=0}^{\infty} \frac{C_n}{R}
\left(\frac{R}{r} \right)^{n+2} P^{(1)}_n(\cos{\theta}) \,, \quad
E_{({\rm azim})} =0 \,, \label{m015}
\end{equation}
do not feel the influence of axions.

\subsection{Axion magnetostatics}

Magnetic field components are defined as follows:
\begin{equation}
B^r \equiv F^{*r0}= - \frac{1}{r^2 \sin{\theta}} \frac{\partial A_{\varphi}}{\partial \theta} = B_{({\rm rad})}  \,,
\label{ms3}
\end{equation}
\begin{equation}
B^{\theta} \equiv F^{*\theta 0}=  \frac{1}{r^2 \sin{\theta}} F_{r \varphi}
\,, \quad B_{({\rm merid})} =  - \frac{1}{r \sin{\theta}} \frac{\partial A_{\varphi}}{\partial r} \,,
\label{ms4}
\end{equation}
\begin{equation}
B^{\varphi} \equiv F^{*\varphi 0}= - \frac{1}{r^2 \sin{\theta}} F_{r \theta}
\,, \quad B_{({\rm azim})}  = -\frac{1}{r} \left[ \frac{\partial A_{\theta}}{\partial r} - \frac{\partial A_{r}}{\partial \theta}  \right] \,,
\label{ms5}
\end{equation}
where the so-called physical components of the magnetic field
\begin{equation}
B_{({\rm rad})} \equiv \sqrt{|B_{r} B^{r}|} \,, \quad B_{({\rm merid})} \equiv \sqrt{|B_{\theta} B^{\theta}|} \,, \quad B_{({\rm azim})} \equiv \sqrt{|B_{\varphi} B^{\varphi}|}
\label{ms6}
\end{equation}
are obtained in analogy with the electric field components (see (\ref{tetrad1})-(\ref{ph})).
Since the equations (\ref{t3}) and (\ref{t4}) convert now into
\begin{equation}
\frac{\partial}{\partial \theta} \left(\sin{\theta} F_{r\theta} - q A_{\varphi}\right) =0 \,,
\quad \frac{\partial}{\partial r} \left(\sin{\theta} F_{r\theta} - q A_{\varphi}\right) =0 \,,
\label{ms8}
\end{equation}
we obtain immediately that
\begin{equation}
F_{r\theta} =  \frac{q}{\sin{\theta}} A_{\varphi} \,.
\label{ms9}
\end{equation}
Then the equation (\ref{t5}) reduces to
\begin{equation}
\left[r^2 \frac{\partial^2}{\partial r^2 } +
\sin{\theta} \frac{\partial}{\partial \theta} \left(\frac{1}{\sin{\theta}}\frac{\partial}{\partial \theta} \right) +
q^2 r^2 \right] A_{\varphi}= 0 \,,
\label{ms11}
\end{equation}
and using the new potential $U(r,\theta)$, defined  by
\begin{equation}
A_{\varphi} = r \sin{\theta} \ U \,, \quad U = \sqrt{|A_{\varphi}A^{\varphi}|} \,,
\label{ms12}
\end{equation}
we obtain the truncated ($\varphi$-independent) Helmholtz equation
\cite{Jackson,book}
\begin{equation}
\Delta_{(1)} U + q^2 U= 0 \,.
\label{ms131}
\end{equation}
The solution of (\ref{ms131}) has the form
\begin{equation}
U = \sum_{n=1}^{\infty} P^{(1)}_{n}(\cos{\theta}) \ \Re_n(r,q) \,,
\label{ms14}
\end{equation}
where the radial function $\Re_n(r,q)$ depends essentially on the
parameter $q$. When $q=0$, i.e., we neglect the axion influence at
all, this function is \cite{Jackson,book}
\begin{equation}
\Re_n(r,0) = {\cal A}_{n} \left(\frac{r}{R}\right)^n + {\cal B}_{n} \left(\frac{r}{R}\right)^{-(n+1)}  \,.
\label{ms141}
\end{equation}
When $q \neq 0$, the radial function can be expressed in terms of
Bessel functions
$$
\Re_n(r,q) = \sqrt{\frac{R}{r}}\left\{ {\cal A}_{n}
\left[ \Gamma\left(n{+}\frac{1}{2} \right) \left(\frac{1}{2}qR\right)^{-\left(n+\frac{1}{2}\right)}\right] J_{n+\frac{1}{2}}(qr)+
\right.
$$
\begin{equation}
\left.
+{\cal B}_{n} \left[(-1)^n \frac{\pi}{\Gamma\left(n{+}\frac{1}{2} \right)} \left(\frac{1}{2}qR \right)^{n+\frac{1}{2}}\right] J_{-\left(n+\frac{1}{2}\right)}(qr) \right\} \,,
\label{ms142}
\end{equation}
where $J_{n+\frac{1}{2}}(qr)$ is the Bessel function of the first
kind with the half-integer index \cite{Jackson,book}. The
constants of integration are chosen so that $\lim_{q \to
0}\Re_n(r,q) = \Re_n(r,0)$ (see (\ref{ms141})).

The corresponding physical components of the  magnetic field are
\begin{equation}
B_{({\rm rad})} = - \sum_{n=1}^{\infty} \frac{n(n{+}1)}{r}  P_n(\cos{\theta}) \ \Re_{n}(r,q) \,,
\label{ms15}
\end{equation}
\begin{equation}
B_{({\rm merid})} = - \sum_{n=1}^{\infty} P^{(1)}_{n}(\cos{\theta}) \frac{1}{r} \frac{d}{dr} \left[r \Re_n(r,q) \right]
 \,,
\label{ms16}
\end{equation}
\begin{equation}
B_{({\rm azim})} =  - q \sum_{n=1}^{\infty} P^{(1)}_{n}(\cos{\theta}) \Re_n(r,q)  \,.
\label{ms17}
\end{equation}
The main new feature is the following: the azimuthal component
$B_{({\rm azim})}$, being equal to zero at $q{=}0$, becomes
nonvanishing at $q\neq 0$.

\subsection{Dipole-type magnetic field}

In the zeroth-order approximation the terrestrial magnetic field
can be described in terms of geomagnetic dipole. In the framework
of this model we obtain basic formulas if put $n=1$ into
(\ref{ms14})-(\ref{ms17}), what yields
\begin{equation}
B_{({\rm rad})} = - \frac{2}{r}  \cos{\theta} \ \Re_{1}(r,q) \,,
\label{ms0159}
\end{equation}
\begin{equation}
B_{({\rm merid})} =   - \frac{\sin{\theta}}{r} \frac{d}{dr} \left[r \Re_1(r,q) \right]
 \,,
\label{ms169}
\end{equation}
\begin{equation}
B_{({\rm azim})} =  - q \sin{\theta} \ \Re_1(r,q)  \,.
\label{ms179}
\end{equation}
We consider two auxiliary angles. The first auxiliary angle
$\gamma$, defined as
\begin{equation}
\tan{\gamma} \equiv \frac{B_{({\rm azim})}(R)}{B_{({\rm rad})}(R)}
= \frac{q R }{2} \tan{\theta} \,, \label{ms19}
\end{equation}
is linear in the parameter $q$ and depends on the meridional  angle $\theta$. As for the second auxiliary angle $\delta$, defined as
\begin{equation}
\tan{\delta} \equiv \frac{B_{({\rm azim})}(R)}{B_{({\rm
merid})}(R)} = q R \left[1+ R \
\frac{\Re^{\prime}_1(R,q)}{\Re_1(R,q)} \right]^{-1}\,,
\label{ms199}
\end{equation}
it does not depend on $\theta$. For the domain $r>R$ by
substitutions  ${\cal A}_{n} =0$ and ${\cal B}_{1} =
\frac{\mu}{R^2}$ we obtain
\begin{equation}
\Re_1(r,q) = \frac{\mu}{r^2} \left(\cos{qr} + qr \sin{qr}\right)\,,
\label{ms09}
\end{equation}
and thus
\begin{equation}
B_{({\rm rad})}(r,q) = - \frac{2 \mu}{r^3} \cos{\theta}
\left(\cos{qr} + qr \sin{qr}\right) \,, \label{ms159}
\end{equation}
\begin{equation}
B_{({\rm merid})}(r,q) =    \frac{\mu \sin{\theta}}{r^3}
\left[\left(\cos{qr} + qr \sin{qr}\right) - q^2 r^2 \cos{qr}
\right]
 \,,
\label{ms1699}
\end{equation}
\begin{equation}
B_{({\rm azim})}(r,q) =  - q  \sin{\theta}\frac{\mu}{r^2}
\left(\cos{qr} + qr \sin{qr}\right) \,. \label{ms917}
\end{equation}
In the approximation $q\to 0$ these formulas give
\begin{equation}
B_{({\rm rad})}(r,0) = - \frac{2 \mu}{r^3} \cos{\theta} \,, \quad
B_{({\rm merid})}(r,0) =    \frac{\mu \sin{\theta}}{r^3}  \,,
\quad B_{({\rm azim})}(r,0) = 0 \,, \label{ms9179}
\end{equation}
as it should be in classical theory of static dipole geomagnetic field. Let us emphasize that in the approximation linear in $q$ the angle $\delta$ is
\begin{equation}
\tan{\delta} =  - q R \,, \label{ms9199}
\end{equation}
thus, the axion-photon coupling can be considered as one of the
explanations of the (visual) magnetic poles drift.

\subsection{Axionically coupled magnetic field as a function of altitude}
\label{altitude}

Classical dipole-type magnetic field decreases monotonically at $r
\to \infty$ according to the law $1/r^3$ (see, e.g.,
(\ref{ms9179})). In the axionic environment, i.e., when $q \neq
0$, the magnetic field components behave non-monotonically (see
(\ref{ms917})). Indeed, there is infinite number of values of the radius  $r{=}R^{*}_{(m)}$, for which the radial function $\Re_1(r,q)$ takes zero value;
they can be found from the equation
\begin{equation}
\cos{[qR^{*}_{(m)}]} + qR^{*}_{(m)} \sin{[qR^{*}_{(m)}]} =0 \,,
\label{strat1}
\end{equation}
or equivalently,
\begin{equation}
{\rm ctg}z = - z \,, \quad z = qR^{*}_{(m)} \,. \label{strat2}
\end{equation}
At the same altitudes $r{=}R^{*}_{(m)}$ the radial and azimuthal
components of the magnetic field change the signs. The position of
the first null is defined by the inequalities $z<\pi$ and
$R^{*}_{(1)}<\pi/ q$. The radial function has infinite number of
extrema at $r=R^{**}_{(j)}$; these quantities satisfy the equation
\begin{equation}
{\rm tg}z = \frac{z}{2} - \frac{1}{z}\,, \quad z = qR^{**}_{(j)}
\,. \label{strat32}
\end{equation}
The first extremum is placed at  $z<3\pi/2$, $R^{**}_{(1)}<3\pi/
2q$. In other words, the relic axions change essentially the
configuration of the magnetic field.

\section{Time dependent electric and magnetic fields in the relic
axion background}

Taking into account the linearity of the model of terrestrial
axion electrodynamics in the relic axion background, we
distinguish two cases in the non-stationary problem. First of all,
we analyze natural oscillations and consider the electric current
four-vector to be vanishing. Then we discuss the effects induced
by the azimuthal electric current.

\subsection{Natural oscillations of electric and magnetic fields in the spherical resonator Earth-Ionosphere}

\subsubsection{Key equations}

When the electric current four-vector is equal to zero, the
equations (\ref{t2})-(\ref{t5}) and (\ref{t14}), (\ref{t144}) can
be reduced to the coupled pair of equations for two new potentials
$U$ and $V$, which play in axion electrodynamics the same role as
the Debye potentials \cite{book} play in the electrodynamics of
Faraday - Maxwell. Let us use the substitutions
\begin{equation}
F_{0r} = - \frac{1}{r \sin{\theta}} \frac{\partial}{\partial \theta}(V \sin{\theta}) = E_{({\rm rad})} \,,
\label{t61}
\end{equation}
\begin{equation}
F_{\theta 0} = -  \frac{\partial}{\partial r} (r V) = -r E_{({\rm merid})}
\,,
\label{t62}
\end{equation}
\begin{equation}
F_{\varphi 0} = -  r \sin{\theta} \frac{\partial}{\partial x^0} U = -  r \sin{\theta} E_{({\rm azim})}
\,,
\label{t623}
\end{equation}
\begin{equation}
F_{r \theta} = r \left(qU +  \frac{\partial}{\partial x^0} V \right)= -r B_{({\rm azim})}
\,,
\label{t624}
\end{equation}
\begin{equation}
F_{r \varphi} = \sin{\theta} \frac{\partial}{\partial r} (r U)= - r \sin{\theta} B_{({\rm merid})}
\,,
\label{t625}
\end{equation}
\begin{equation}
F_{\theta \varphi} = r \frac{\partial}{\partial \theta} (U \sin{\theta}) = - r^2 \sin{\theta} B_{({\rm rad})}
\,,
\label{t626}
\end{equation}
\begin{equation}
A_{\varphi} =  r U \sin{\theta} \,.
 \label{t63}
\end{equation}
Then the equations (\ref{t2}), (\ref{t3}), (\ref{t4}) and
(\ref{t144}) turn into identities, the equation (\ref{t14}) gives
\begin{equation}
\Delta_{(1)} V - \frac{\partial^2}{\partial x^{0 2}} V = q
\frac{\partial}{\partial x^{0}}U \,, \label{t17}
\end{equation}
and the equations (\ref{t5}) takes the form
\begin{equation}
\Delta_{(1)} U - \frac{\partial^2}{\partial x^{0 2}} U = - q^2 U  - q \frac{\partial}{\partial x^{0}}V \,.
\label{t16}
\end{equation}
We indicate the pair of equations (\ref{t17}) and (\ref{t16}) as
key equations. When $q \neq 0$, using the decoupling procedure for
the equations (\ref{t17}) and (\ref{t16}), we obtain the key
equation of the fourth order for the function $V$:
\begin{equation}
\left\{\left[\Delta_{(1)}  - \frac{\partial^2}{\partial x^{0 2}}\right]^2  + q^2 \Delta_{(1)} \right\} V =0 \,.
\label{t18}
\end{equation}
When $V$ is found, the function $U$ can be extracted from
(\ref{t17}) by integration over time.

\subsubsection{Boundary value problem, eigenvalues and eigenfunctions}

Let us consider electromagnetic oscillations in the spherical
resonator bounded by the Earth surface ($r{=}R$) and the bottom
edge of the Earth Ionosphere ($r{=}R_{*}$) in the framework of the
truncated model.

Let the quantities ${\cal F}_{nj}(r,\theta)$ be the eigenfunctions
of the operator $\Delta_{(1)}$ satisfying the boundary conditions
\begin{equation}
{\cal F}_{nj}(R,\theta) = 0 \,, \quad {\cal F}_{nj}(R_{*},\theta)
= 0 \,. \label{eig4}
\end{equation}
They have the well-known multiplicative form
\begin{equation}
{\cal F}_{nj}(r,\theta) = P^{(1)}_{n}(\cos{\theta}) \ {\cal
H}_{nj}(r) \,, \label{eig1}
\end{equation}
where the radial functions ${\cal H}_{nj}(r)$ are
\begin{equation}
{\cal H}_{nj}(r) {=} \frac{1}{\sqrt{r}}\left[
J_{n{+}\frac{1}{2}}\left(\nu^{(n)}_{j} r \right)
J_{{-}\left(n{+}\frac{1}{2}\right)}\left(\nu^{(n)}_{j} R \right)
{-} J_{n{+}\frac{1}{2}}\left(\nu^{(n)}_{j} R \right)
J_{{-}\left(n{+}\frac{1}{2}\right)}\left(\nu^{(n)}_{j} r \right)
\right] \,.\label{eig71}
\end{equation}
Boundary conditions (\ref{eig4}) are satisfied, when the
parameters $\nu^{(n)}_j$ are found from the equation
\begin{equation}
J_{n+\frac{1}{2}}\left(\nu^{(n)}_{j} R \right)  J_{-\left(n+\frac{1}{2}\right)}\left(\nu^{(n)}_{j} R_{*}\right)
= J_{n+\frac{1}{2}}\left(\nu^{(n)}_{j} R_{*}\right) J_{-\left(n+\frac{1}{2}\right)}\left(\nu^{(n)}_{j} R\right)
\,,
\label{eig5}
\end{equation}
where the index $j=0,1,2,...$ counts the positive zeros of the
equations  $(\ref{eig5})$. Of course, the boundary conditions
(\ref{eig4}) are chosen as the simplest example, which can
illustrate our main statements; as for applications, we intend to
use more sophisticated boundary conditions in special papers.

We search for the potential functions $U$ and $V$ in the form
\begin{equation}
U(\tilde{t},r,\theta) = \sum_{n=0}^{\infty} \sum_{j=0}^{\infty}
u_{nj}(\tilde{t}) \ P^{(1)}_n(\cos{\theta}) \ {\cal H}_{nj}(r) \,,
\label{eig6}
\end{equation}
\begin{equation}
V(\tilde{t},r,\theta) = \sum_{n=0}^{\infty} \sum_{j=0}^{\infty}
v_{nj}(\tilde{t}) \ P^{(1)}_n(\cos{\theta}) \ {\cal H}_{nj}(r) \,,
\label{eig7}
\end{equation}
where we put for the sake of simplicity that $x^0 \equiv c
\tilde{t}$. The key equations (\ref{t16}) and (\ref{t17}) give the
following coupled pair of equations for the time dependent
amplitudes:
\begin{equation}
\ddot{u}_{nj} + c^2 \left[\left(\nu^{(n)}_{j}\right)^2 - q^2 \right]u_{nj} - qc \dot{v}_{nj} = 0
\,,
\label{eig8}
\end{equation}
\begin{equation}
\ddot{v}_{nj} + c^2 \left(\nu^{(n)}_{j}\right)^2 v_{nj} + qc \dot{u}_{nj} = 0
\,.
\label{eig9}
\end{equation}
The axionic parameter enters these equation non-equivalently.

\subsubsection{Solutions for the mode amplitudes}

In order to simplify the analysis we suppose here that the axionic
parameter $q$ is positive, $q>0$. The system of coupled equations
(\ref{eig8}) and (\ref{eig9}) has standardly the solutions of
three types. When $q$ is so small that $q<\nu^{(n)}_{j}$ for all
$j$ and $n$, we obtain harmonic oscillations for both functions
\begin{equation}
v_{nj}(\tilde{t}) = \left[C^{(1)}_{1nj}
\cos{\omega_{1nj}\tilde{t}} + C^{(1)}_{2nj}
\sin{\omega_{1nj}\tilde{t}} \right] + \left[C^{(2)}_{1nj}
\cos{\omega_{2nj}\tilde{t}} + C^{(2)}_{2nj}
\sin{\omega_{2nj}\tilde{t}} \right] \,, \label{eig10}
\end{equation}
and
$$
u_{nj}(\tilde{t}) {=}
\frac{1}{\sqrt{1{-}\frac{q}{\nu^{(n)}_{j}}}}\left[C^{(1)}_{2nj}
\cos{\omega_{1nj}\tilde{t}} {-} C^{(1)}_{1nj}
\sin{\omega_{1nj}\tilde{t}} \right] {+}
$$
\begin{equation}
{+}\frac{1}{\sqrt{1{+}\frac{q}{\nu^{(n)}_{j}}}}
\left[C^{(2)}_{1nj} \sin{\omega_{2nj}\tilde{t}} {-} C^{(2)}_{2nj}
\cos{\omega_{2nj}\tilde{t}} \right] \,. \label{eig11}
\end{equation}
The constants $C^{(1)}_{1nj}$, $C^{(1)}_{2nj}$, $C^{(2)}_{1nj}$
and $C^{(1)}_{2nj}$ can be found from initial data. The hybrid
frequencies
\begin{equation}
\omega_{1nj} = c \nu^{(n)}_{j}\sqrt{1{-}\frac{q}{\nu^{(n)}_{j}}} \,, \quad
\omega_{2nj} = c \nu^{(n)}_{j}\sqrt{1{+}\frac{q}{\nu^{(n)}_{j}}}
\,,
\label{eig12}
\end{equation}
coincide when $q{=}0$, i.e., when there is no axion-photon
coupling.

Formally speaking, for some mode numbers, say, $n<n_{*}$ and
$j<j_{*}$, the quantity $q$ can exceed $\nu^{(n_{*})}_{j_{*}}$.
The corresponding modes are non-harmonic and one should replace
the first pair $\cos{\omega_{1nj}\tilde{t}}$ and
$\sin{\omega_{1nj}\tilde{t}}$ by hyperbolic functions
$\cosh{\Gamma_{1nj}\tilde{t}}$ and $\sinh{\Gamma_{1nj}\tilde{t}}$,
where $\Gamma_{1nj} \equiv c
\nu^{(n)}_{j}\sqrt{\frac{q}{\nu^{(n)}_{j}}{-}1}$.

When $q {=}\nu^{(n_{*})}_{j_{*}}$, the corresponding mode
amplitudes
\begin{equation}
v_{*}(\tilde{t}) =
\frac{1}{2}\left[v_{*}(0){-}\frac{\dot{u}_{*}(0)}{cq} \right] +
\frac{1}{2}\left[v_{*}(0){+}\frac{\dot{u}_{*}(0)}{cq} \right]
\cos{\sqrt{2} cq \tilde{t}} + \frac{\dot{v}_{*}(0)}{\sqrt{2} cq}
\sin{\sqrt{2} cq \tilde{t}} \,, \label{res1}
\end{equation}
$$
u_{*}(\tilde{t}) = \left[u_{*}(0){+}\frac{\dot{v}_{*}(0)}{2cq}
\right] -
\frac{cq\tilde{t}}{2}\left[v_{*}(0){-}\frac{\dot{u}_{*}(0)}{cq}
\right] +
$$
\begin{equation}
+ \frac{1}{2\sqrt2}\left[v_{*}(0){+}\frac{\dot{u}_{*}(0)}{cq}
\right]\sin{\sqrt{2} cq\tilde{t}} - \frac{\dot{v}_{*}(0)}{2cq}
\cos{\sqrt{2} cq\tilde{t}} \,, \label{res2}
\end{equation}
display the possibility of oscillations on the specific (axionic)
frequency $\omega_{{\rm A}}{=}\sqrt2 cq$, as well as, the linear
growth with time of the $U$-potential.

\subsection{Forced oscillations produced by an azimuthal electric current}

Let us consider the models in which the only azimuthal component
of the electric current four-vector $J^{\varphi}$ is
non-vanishing. This case is very illustrative just for the
truncated ($\varphi$-independent) model. Indeed, the charge
conservation law (\ref{t51}) is now satisfied identically. The
electrodynamic equations (\ref{t17}), (\ref{t16}) with the
definitions (\ref{t61})-(\ref{t63}) transform into
\begin{equation}
\Delta_{(1)} V - \frac{\partial^2}{\partial x^{0 2}} V = q \frac{\partial}{\partial x^{0}}U \,,
\label{0t17}
\end{equation}
\begin{equation}
\Delta_{(1)} U - \frac{\partial^2}{\partial x^{0 2}} U = - q^2 U  - q \frac{\partial}{\partial x^{0}}V + \frac{4\pi}{c} I \,,
\label{0t16}
\end{equation}
where
\begin{equation}
I \equiv \sqrt{|I^{\varphi} I_{\varphi}|} = I^{\varphi} r
\sin{\theta} \,. \label{00t18}
\end{equation}
Decoupled equation for $V$ is modified accordingly as
\begin{equation}
\left\{\left[\Delta_{(1)}  - \frac{\partial^2}{\partial x^{0 2}}\right]^2  + q^2 \Delta_{(1)} \right\} V = \frac{4\pi q}{c} \frac{\partial}{\partial x^{0}} I  \,.
\label{0t18}
\end{equation}
Repeating the decomposition of the potentials $V$ and $U$
fulfilled in Subsection 5.1., we obtain the differential equations
\begin{equation}
\ddot{u}_{nj} + c^2 \left[\left(\nu^{(n)}_{j}\right)^2 - q^2 \right]u_{nj} - qc \dot{v}_{nj} = I_{nj}(\tilde{t})
\,,
\label{eig80}
\end{equation}
\begin{equation}
\ddot{v}_{nj} + c^2 \left(\nu^{(n)}_{j}\right)^2 v_{nj} + qc \dot{u}_{nj} = 0
\,,
\label{eig90}
\end{equation}
where the functions $I_{nj}(\tilde{t})$ denote the coefficients in the decomposition
\begin{equation}
\frac{4\pi}{c} I(\tilde{t},r,\theta) = \sum_{n=0}^{\infty}
\sum_{j=0}^{\infty} I_{nj}(\tilde{t}) P^{(1)}_n(\cos{\theta}){\cal
H}_{nj}(r) \,. \label{0eig6}
\end{equation}
When $q{=0}$, the key equations are not coupled, so the only $U$
potential feels the influence of the stimulating current of this
type. The case $q \neq 0$ is much more sophisticated.

The particular solutions $v^{*}_{nj}(\tilde{t})$, which satisfy
(\ref{eig80}) and (\ref{eig90}) and null initial data
$v^{*}_{nj}(0)=0=\dot{v}^{*}_{nj}(0)$, have the following form
\begin{equation}
v^{*}_{nj}(\tilde{t}) = \frac{1}{\sqrt2
\sqrt{\omega^2_{1nj}{+}\omega^2_{2nj}}} \int_0^{\tilde{t}} d\tau
I_{nj}(\tau) \left[\cos{\omega_{2nj}(\tilde{t}{-}\tau)}-
\cos{\omega_{1nj}(\tilde{t}{-}\tau)} \right] \,. \label{0eig10}
\end{equation}
Clearly, this solution vanishes at $q \to 0$, since in this limit $\omega_{2nj} \to \omega_{1nj}$.

The particular solutions $u^{*}_{nj}(\tilde{t})$ with null initial
data $u^{*}_{nj}(0) = 0 = \dot{u}^{*}_{nj}(0)$ is
\begin{equation}
u^{*}_{nj}(\tilde{t}) = \frac{1}{2} \int_0^{\tilde{t}} d\tau
I_{nj}(\tau) \left[
\frac{\sin{\omega_{1nj}(\tilde{t}{-}\tau)}}{\omega_{1nj}}+
\frac{\sin{\omega_{2nj}(\tilde{t}{-}\tau)}}{\omega_{2nj}} \right]
\,. \label{0eig11}
\end{equation}
Thus, when $q{=}0$, the azimuthal current produces stimulated
azimuthal electric field, meridional and radial magnetic field.
When $q \neq 0$, additional radial and meridional electric field,
as well as the azimuthal magnetic field also appears as a result
of the axion-photon coupling.
When the azimuthal component of the electric current is periodic,
and the frequency is, say, $\Omega$, i.e.,
\begin{equation}
I_{nj}(\tilde{t}) = I_{nj} \ \sin{\Omega \tilde{t}} \,, \label{I2}
\end{equation}
we obtain stimulated oscillations on the heterodyne frequencies
$\omega_{1nj}\pm \Omega$ and $\omega_{2nj}\pm \Omega$. For
instance, for the amplitude $u^{*}_{nj}(\tilde{t})$ one obtains:
$$
u^{*}_{nj}(\tilde{t}) = \frac{1}{4 \omega^2_{1nj}\omega^2_{2nj}}
I_{nj} \left\{ 2 \left(\omega^2_{1nj}{+}\omega^2_{2nj}
\right)\sin{\Omega \tilde{t}} {+} \omega^2_{2nj}\left[
\sin{(\omega_{1nj}{-}\Omega)\tilde{t}} {-}
\sin{(\omega_{1nj}{+}\Omega)\tilde{t}} \right] {+} \right.
$$
\begin{equation}
\left. + \omega^2_{1nj}\left[
\sin{(\omega_{2nj}{-}\Omega)\tilde{t}} -
\sin{(\omega_{2nj}{+}\Omega)\tilde{t}} \right] \right\}\,.
\label{period1}
\end{equation}
For the mode amplitude  $v^{*}_{nj}(\tilde{t})$ the result is
similar.

\section{An application: axionically coupled electric and magnetic fields near the Earth's
surface}

\subsection{Longitudinal cluster formed by meridional electric and magnetic fields}

Let us consider the electric and magnetic fields at the Earth
surface $r=R$, using the definitions (\ref{t61})-(\ref{t626}) and
the solutions (\ref{eig6})-(\ref{eig12}) to the key equations. The
radial and azimuthal components of the electric field in this
model vanish at the Earth surface
\begin{equation}
E_{({\rm rad})}(\tilde{t},R,\theta) = 0 \,, \quad E_{({\rm
azim})}(\tilde{t},R,\theta) = 0 \,. \label{eig13}
\end{equation}
The radial and azimuthal components of the magnetic field also
take zero values
\begin{equation} B_{({\rm
rad})}(\tilde{t},R,\theta) = 0 \,, \quad B_{({\rm
azim})}(\tilde{t},R,\theta) = 0 \,. \label{eig14}
\end{equation}
Only meridional components of electric field
\begin{equation}
E_{({\rm merid})}(\tilde{t},R,\theta) = \sum_{n=0}^{\infty}
\sum_{j=0}^{\infty} v_{nj}(\tilde{t}) \ P^{(1)}_n(\cos{\theta}) \
{\cal G}_{nj}(R) \,, \label{eig15}
\end{equation}
and magnetic field
\begin{equation}
B_{({\rm merid})}(\tilde{t},R,\theta) = - \sum_{n=0}^{\infty}
\sum_{j=0}^{\infty} u_{nj}(\tilde{t}) \ P^{(1)}_n(\cos{\theta}) \
{\cal G}_{nj}(R) \,, \label{eig16}
\end{equation}
are non-vanishing and thus attract attention in this model. In the
formulas (\ref{eig15}) and (\ref{eig16}) we used the auxiliary
function
\begin{equation}
{\cal G}_{nj}(R) {=} \frac{\nu^{(n)}_j}{\sqrt{R}}\left[J^{\prime}_{n{+}
\frac{1}{2}}\left(\nu^{(n)}_{j} R \right)  J_{{-}\left(n{+}\frac{1}{2}\right)}\left(\nu^{(n)}_{j} R \right)
{-} J_{n{+}\frac{1}{2}}\left(\nu^{(n)}_{j} R \right) J^{\prime}_{{-}\left(n{+}\frac{1}{2}\right)}\left(\nu^{(n)}_{j} R \right) \right]
\,.
\label{eig17}
\end{equation}
Clearly, the meridional electric and magnetic fields are coupled
by the axion field, and we deal with longitudinal cluster of
electromagnetic modes. Let us analyze the time dependent
variations of these fields by two examples.

\subsection{Perturbations of meridional electric field produce
variations of the meridional magnetic field}

Let us consider, first, the situation when initial data are the following
\begin{equation}
v_{nj}(0) \neq 0 \,, \quad \dot{v}_{nj}(0) \neq 0 \,, \quad u_{nj}(0)=0 \,, \quad \dot{u}_{nj}(0) =0
\,.
\label{0eig18}
\end{equation}
This means that initially only the meridional electric field is
perturbed. The corresponding solution for the $v_{nj}$ amplitude
is (see (\ref{eig10}))
$$
v_{nj}(\tilde{t}) = \frac{1}{2}v_{nj}(0)\left[\cos{\omega_{1nj}\tilde{t}} + \cos{\omega_{2nj}\tilde{t}} \right] +
$$
\begin{equation}
+\frac{1}{2}\dot{v}_{nj}(0) \frac{1}{\omega_{1nj}\omega_{2nj}}
\gamma^2_{(n)j}\left[\omega_{1nj}\sin{\omega_{1nj}t} +
\omega_{2nj}\sin{\omega_{2nj}t} \right] \,.
\label{eig18}
\end{equation}
The solution for the $u_{nj}$ amplitude gives now non-vanishing
result, though these functions started from zero initial values
(see (\ref{0eig18})):
$$
u_{nj}(\tilde{t}) = \frac{1}{2\sqrt{\omega_{1nj}\omega_{2nj}}} \gamma^{-1}_{(n)j}
\left\{v_{nj}(0) \left[\omega_{1nj}\sin{\omega_{2nj}\tilde{t}} - \omega_{2nj}\sin{\omega_{1nj}\tilde{t}} \right]
+
\right.
$$
\begin{equation}
\left. +
\dot{v}_{nj}(0)\gamma^2_{(n)j}\left[\cos{\omega_{1nj}\tilde{t}} -
\cos{\omega_{2nj}\tilde{t}} \right] \right\} \,.
\label{eig19}
\end{equation}
Here we introduced the following convenient constants:
\begin{equation}
\gamma_{(n)j} \equiv \left[1{-}q^2 \left(\nu^{(n)}_{j}\right)^{-2} \right]^{-\frac{1}{4}} \,.
\label{eig191}
\end{equation}
For the resonant case, when $q{=}\nu^{n_{*}}_{j_{*}}$ for some
mode numbers $n_{*}$ and $j_{*}$, one can illustrate the $U$-mode
amplitude generation by the formula
\begin{equation}
u_{*}(\tilde{t})= \frac{v_{*}(0)}{2\sqrt2}\left[\sin{\sqrt2cq\tilde{t}} - \sqrt2cq\tilde{t} \right]
+ \frac{\dot{v}_{*}(0)}{cq}\sin^2{\frac{cq\tilde{t}}{\sqrt2}} \,.
\label{res3}
\end{equation}
Clearly, there is the term linear in time, and for small value of $cq\tilde{t}$ the expression (\ref{res3}) has the leading order term $\propto cq\tilde{t}^2$.

If $q=0$, so that  $\omega_{1nj}{=}\omega_{2nj} \equiv \omega_{0nj}$, one obtains that $u_{nj}(t)=0$, thus the magnetic field remains unperturbed.
In the approximation linear with respect to $q$ one obtains
\begin{equation}
v_{nj}(\tilde{t}) = v_{nj}(0)\cos{\omega_{0nj}\tilde{t}} + \frac{\dot{v}_{nj}(0)}{\omega_{0nj}}\sin{\omega_{0nj}\tilde{t}} \,,
\label{eig25}
\end{equation}
\begin{equation}
u_{nj}(\tilde{t}) = \frac{qc}{2\omega_{0nj}} \left\{v_{nj}(0)\left[\omega_{0nj} \tilde{t} \cos{\omega_{0nj}\tilde{t}} - \sin{\omega_{0nj}\tilde{t}} \right] + \dot{v}_{nj}(0) \tilde{t} \sin{\omega_{0nj}\tilde{t}} \right\}\,.
\label{eig26}
\end{equation}
In particular, we obtain very convenient formula for estimations
\begin{equation}
u_{nj}(\tilde{t}) = \frac{1}{2}q c \tilde{t}  \ v_{nj}(\tilde{t})  \,,
\label{eig27}
\end{equation}
when the initial data have specific form $v_{nj}(0)=0$.

\subsection{Perturbations of meridional magnetic field produce
variations of the meridional electric field}

Let us consider now the situation when initial data are the following
\begin{equation}
u_{nj}(0) \neq 0 \,, \quad \dot{u}_{nj}(0) \neq 0 \,, \quad
v_{nj}(0)=0 \,, \quad \dot{v}_{nj}(0) =0 \,. \label{0eig186}
\end{equation}
This means that initially  only the meridional magnetic field is
perturbed. Now the $v_{nj}$ amplitudes become non-vanishing at
$\tilde{t}>0$ due to the axion-photon coupling:
$$
v_{nj}(\tilde{t}) =
\frac{\gamma_{(n)j}}{2\sqrt{\omega_{1nj}\omega_{2nj}}} \left\{
\dot{u}_{nj}(0)\left[\cos{\omega_{2nj}\tilde{t}} -
\cos{\omega_{1nj}\tilde{t}} \right] + \right.
$$
\begin{equation}
\left. + u_{nj}(0)\gamma^2_{(n)j}
\left[\omega_{2nj}\sin{\omega_{1nj}\tilde{t}} -
\omega_{1nj}\sin{\omega_{2nj}\tilde{t}} \right] \right\} \,.
\label{eig22}
\end{equation}
The $u_{nj}$ amplitudes evolve as follows:
$$
u_{nj}(\tilde{t}) =
\frac{1}{2 \omega_{1nj}\omega_{2nj}} \left\{
u_{nj}(0) \gamma^2_{(n)j}\left[\omega^2_{2nj} \cos{\omega_{1nj}\tilde{t}} + \omega^2_{1nj}\cos{\omega_{2nj}\tilde{t}} \right]
+
\right.
$$
\begin{equation}
\left.
+ \dot{u}_{nj}(0)\left[\omega_{2nj}\sin{\omega_{1nj}\tilde{t}} + \omega_{1nj}\sin{\omega_{2nj}\tilde{t}} \right]\right\}
\,.
\label{eig23}
\end{equation}
Again, when $q{=}\nu^{n_{*}}_{j_{*}}$, one can illustrate the $v$-mode amplitude generation by the formulas
\begin{equation}
v_{*}(\tilde{t})= - \frac{\dot{u}_{*}(0)}{cq}
\sin^2{\frac{cq\tilde{t}}{\sqrt2}} \,, \quad u_{*}(t)= u_{*}(0) +
\frac{\dot{u}_{*}(0)}{2\sqrt2 cq}\left[\sin{\sqrt2 cq\tilde{t}} +
\sqrt2 cq\tilde{t} \right] \,. \label{res31}
\end{equation}
The term linear in time appears now in $u_{*}(t)$.

Clearly, if $q{=}0$, so that  $\omega_{1nj}{=}\omega_{2nj}$, one obtains that $v_{nj}(\tilde{t}){=}0$, thus the electric field remains unperturbed.
Analogously, in the linear approximation we obtain
\begin{equation}
u_{nj}(\tilde{t}) = u_{nj}(0)\cos{\omega_{0nj}\tilde{t}} + \frac{\dot{v}_{nj}(0)}{\omega_{0nj}}\sin{\omega_{0nj}\tilde{t}} \,,
\label{eig259}
\end{equation}
\begin{equation}
v_{nj}(\tilde{t}) = - \frac{qc}{2\omega_{0nj}} \left\{u_{nj}(0)\left[\omega_{0nj} \tilde{t} \cos{\omega_{0nj}\tilde{t}} - \sin{\omega_{0nj}\tilde{t}} \right] + \dot{u}_{nj}(0) t \sin{\omega_{0nj}\tilde{t}} \right\}\,,
\label{eig269}
\end{equation}
with
\begin{equation}
v_{nj}(\tilde{t}) = - \frac{1}{2}q c \tilde{t} \ u_{nj}(\tilde{t})
\,, \label{eig279}
\end{equation}
when the initial data are chosen as $u_{nj}(0)=0$.

\section{Discussion}

1. We studied one model of coupling of terrestrial magnetic and
electric fields with a relic axion background. Exact solutions
obtained in the framework of the axion electrodynamics illustrate
three important conclusions.

1.1. Relic dark matter axions produce in the terrestrial
electrodynamic system oscillations of a new type, which belong to
the class of Longitudinal Electro-Magnetic Clusters (LEMCs). LEMC
has the following specific feature: the electric and magnetic
fields are parallel to one another and are coupled by axionic
field (see, e.g., the formulas (\ref{eig13})-(\ref{eig17})); in
the absence of axions such oscillations decouple. Electric and
magnetic fields of this type are correlated: the oscillations are
characterized by identical frequencies but different phases (see,
e.g., the formulas (\ref{eig18})-(\ref{eig19})) and
(\ref{eig22})-(\ref{eig23}))). Generally, there are two sets of
hybrid frequencies of LEMCs (see the formulas (\ref{eig12})), and
these frequencies coincide when the axionic dark matter influence
is negligible.

1.2. Relic dark matter axions  deform the static terrestrial
magnetic field. First, in case when the original geomagnetic field
has the radial and meridional components only, the axion-photon
coupling produces a supplementary azimuthal component (see, e.g.,
(\ref{ms17}), (\ref{ms917})); this effect contributes to the
phenomenon of the Earth's magnetic pole drift. Second, due to the
axion-photon coupling the dependence of the magnetic field on the
altitude becomes non-monotonic: the surfaces appear, on which the
radial and/or meridional components of the geomagnetic field
change the signs, as well as reach maxima and minima (see
Subsection \ref{altitude}).

1.3. The truncated ($\varphi$-independent) model can be
effectively described in terms of two electromagnetic potentials
$U$ and $V$, which can be considered as  axionic
generalizations of the well-known Debye potentials.

2. Estimations of the effect produced by relic axion background in
the terrestrial electrodynamic system depend on how small the
dimensionless parameter $\xi \equiv q R$ is. It can be estimated
as follows. Let us remind that in the laboratory frame of
reference $|q| {=} \frac{c}{\Psi_0} \sqrt{\rho_{({\rm DM})}}$, and
use the natural units, in which $\hbar{=}1$ and $c{=1}$. According
to \cite{DM99} the mass density of the dark matter in the Solar
system is estimated to be $\rho_{({\rm DM})} \simeq 0.033 \
M_{({\rm Sun})} {\rm pc}^{-3}$ or in the natural units $\rho_{({\rm
DM})} \simeq 1.25 \ {\rm GeV} \cdot {\rm cm}^{-3}$. According to \cite{10,n1}
the parameter $\Psi_0$ is reciprocal to the axion-photon-photon
coupling constant $\rho_{{\rm A} \gamma \gamma}$, i.e., $
\frac{1}{\Psi_0}{=} \rho_{{\rm A} \gamma \gamma}$ and $\rho_{{\rm
A} \gamma \gamma}$ itself can be (optimistically) estimated as
$\rho_{{\rm A} \gamma \gamma} \simeq 10^{-9} {\rm GeV}^{-1}$. Thus, we
obtain $\xi \simeq 10^{-7}$. This means that the effective
frequency of LEMCs is about $\nu_{{\rm A}} {=} cq {=} \frac{c
\xi}{R} \simeq 0.5 \cdot 10^{-5} {\rm Hz}$. This frequency belongs to
the range of infra-low frequencies; variations of electric  and
magnetic fields of this type are studying in the experimental
group of Vladimir University during more than 40 years
\cite{Vlad1,Vlad2,Vlad3}.

3. We consider this paper as theoretical grounds for the program
of testing of the LEMCs in the Earth's atmosphere. Detailed
technical description of the experiments and first experimental
results are planned to be discussed in a special paper.

\vspace{5mm} \noindent {\large\bf Acknowledgments}

\noindent This work was supported by the FTP "Scientific and
Scientific - Pedagogical Personnel of the Innovative Russia"
(grants Nos 16.740.11.0185 and 14.740.11.0407), and by the Russian
Foundation for Basic Research (grant Nos. 11-02-01162 and
11-05-97518 - p-center-a).

\end{document}